\documentstyle[a4,12pt,epsf,amsfonts,amssymb]{article}
\begin{document} 

\title{Teaching General Relativity}  

\author{Robert M. Wald \\ 
 \it Enrico Fermi Institute and Department of Physics \\  
 \it The University of Chicago, Chicago, IL 60637, USA} 

\maketitle

\begin{abstract}

\end{abstract}

This Resource Letter provides some guidance on issues that arise in
teaching general relativity at both the undergraduate and graduate
levels. Particular emphasis is placed on strategies for presenting the
mathematical material needed for the formulation of general
relativity.

\section{Introduction}

General Relativity is the theory of space, time, and gravity
formulated by Einstein in 1915. It is widely regarded as a very
abstruse, mathematical theory and, indeed, until recently it has not
generally been regarded as a suitable subject for an undergraduate
course.  In actual fact, the the mathematical material (namely,
differential geometry) needed to attain a deep understanding of
general relativity is not particularly difficult and requires a
background no greater than that provided by standard courses in
advanced calculus and linear algebra. (By contrast, considerably more
mathematical sophistication is need to provide a rigorous formulation
of quantum theory.)  Nevertheless, this mathematical material is
unfamiliar to most physics students and its application to general
relativity goes against what students have been taught since high
school (or earlier): namely, that ``space'' has the natural structure of a
vector space.  Thus, the mathematical material poses a major challenge
to teaching general relativity---particularly for a one-semester
course.  If one take the time to teach the mathematical material
properly, one runs the risk of turning the course into a course on
differential geometry and doing very little physics. On the other
hand, if one does not teach it properly, then one is greatly
handicapped in one's ability to explain the major conceptual
differences between general relativity and the pre-relativitistic and
special relativistic notions of spacetime structure.

The purpose of this Resource Letter is to provide a brief guide to the
issues and pitfalls involved in teaching general relativity at both
the undergraduate and graduate level. The main focus will be on how to
introduce the mathematical material necessary for the formulation of
general relativity. By contrast, I shall not devote much attention to
how to teach the various topics that normally would be included in a
general relativity course after one has formulated the theory, such as
the ``weak field'' limit, tests of general relativity, gravitational
radiation, cosmology, and black holes.  This Resource Letter also will be
relatively light on the enumeration of ``resources''.

I will begin by briefly outlining the major new conceptual ideas
introduced by general relativity.  I will then describe the
mathematical concepts that are needed to formulate the theory in a
precise manner. Finally, I will discuss strategies for dealing with
this mathematical material in courses on general relativity.

\section{General Relativity}

Prior to 1905, it was taken for granted that the causal structure of
spacetime defines a notion of {\it simultaneity}. For a given event
$A$ (i.e., a ``point of space at an instant of time''), we can define
the {\em future of $A$} to consist of all events that, in principle,
could be reached by a particle starting from event $A$. Similarly, the
past of $A$ consists of all events such that, in principle, a particle
starting from that event could arrive at $A$. The events that lie
neither to the future nor the past of event $A$ were assumed to
comprise a $3$-dimensional set, called the events {\em simultaneous}
with $A$. This notion of simultaneity defines a notion of ``all of
space at an instant of time'', which, in essence, allows one to
decompose the study of spacetime into separate studies of ``space''
and ``time''. It is important to emphasize to students the key role of
this assumption in pre-relativistic notions of spacetime structure.

The major revolution introduced by special relativity is largely
premised on the fact that the assertions of the previous paragraph
concerning the causal structure of spacetime are wrong. Most
strikingly, the set of events that fail to be causally connected to an
event $A$ comprise much more than a $3$-dimensional region. In a
spacetime diagram, the future of an event $A$ looks like the interior
of a ``cone'' with vertex $A$, where the boundary of this cone
corresponds to the trajectories of light rays emitted at event $A$.
Thus, in special relativity, the causal structure of spacetime defines
a notion of a ``light cone'' of an event, but it does not define a
notion of simultaneity.

It is important to focus on the ``invariant structure'' of spacetime,
i.e., the aspects of spacetime structure that are well defined,
independently of which observer makes the measurements. In
pre-relativity physics, the time interval between any pair of events
is such an invariant; the space interval between simultaneous events
is also an invariant. However, in special relativity neither time
intervals nor space intervals are invariants. In special relativity,
the only invariant quantity related to a pair of events, $A$ and $B$,
is their {\it spacetime interval}, given in any global inertial
coordinate system by the formula 
\begin{equation} 
I(A,B) = -(\Delta t)^2 + \frac{1}{c^2}[(\Delta x)^2 
+(\Delta y)^2 + (\Delta z)^2]
\label{int}
\end{equation} 
All features of spacetime structure in special
relativity can be derived from the spacetime interval.

It is a remarkable fact that---except for the key minus sign in front
of $(\Delta t)^2$---the spacetime interval has exactly the same
mathematical form as the Pythagorean formula for the square of the
distance between two points in Euclidean geometry. The fact was first
realized by Minkowski in 1908, but its deep significance was not
appreciated by Einstein until several years later, as he began to
develop general relativity. It enables one to understand special
relativity as a theory of {\it flat Lorentzian geometry}. In special
relativity, spacetime is described in a manner which is mathematically
identical to Euclidean geometry, except for the changes that result
from the presence of a term with a minus sign on the right side of
eq.(\ref{int}). In particular, the global inertial coordinates of
special relativity are direct analogs of Cartesian coordinates in
Euclidean geometry, and the worldlines of inertial observers are
direct analogs of the straight lines (geodesics) of Euclidean geometry.

This understanding of special relativity as a theory of flat
Lorentzian geometry is a key step in the progression towards general
relativity. General relativity arose from the attempt to formulate a
theory of gravity that is compatible with the basic ideas of special
relativity and also fundamentally builds in the {\it equivalence
principle}: All bodies are affected by gravity and, indeed, all bodies
fall the same way in a gravitational field. The equivalence principle
strongly suggests that freely falling motion in a gravitational field
should be viewed as analogous to inertial motion in pre-relativity
physics and special relativity. Gravity isn't a ``force'' at all, but
rather a change in spacetime structure that allows inertial observers
to accelerate relative to each other. Remarkably, after many years of
effort, Einstein discovered that this idea could be implemented by
simply generalizing the flat Lorentzian geometry of special relativity to a
curved Lorentzian geometry---in exactly the same way as flat Euclidean geometry
can be generalized to curved Riemannian geometry. General relativity
is thereby a theory of the structure of space and time that accounts
for all of the physical effects of gravitation in terms of the curved
geometry of spacetime.

In addition to the replacement of a flat spacetime geometry by a
curved spacetime geometry, general relativity differs radically from
special relativity in that the spacetime geometry is not fixed in
advance but rather evolves dynamically. The dynamical evolution
equation for the metric---known as {\it Einstein's equation}---equates
part of the curvature of spacetime to the stress-energy-momentum
tensor of matter.

\section{Differential Geometry}

The geometry required for an understanding of general relativity is
simply the generalization of Riemannian geometry to metrics that are
not positive-definite. Fortunately, there are few significant
mathematical changes that result from this generalization.
Consequently, much of the intuition that most people have for
understanding the Riemannian geometry of two-dimensional surfaces
encountered in everyday life---such as the surface of a potato---can
usually be extended to general relativity in a reliable
manner. However, two significant cautions should be kept in mind: (1)
Much of the intuition that most people have about the curvature of
two-dimensional surfaces concerns the manner in which the surface {\it
bends} within the three-dimensional Euclidean space in which it
lies. This {\it extrinsic} notion of curvature must be carefully
distinguished from the purely {\it intrinsic} notion of curvature that
concerns, e.g., the failure of initially parallel geodesics within the
surface itself to remain parallel. It is the intrinsic notion of curvature
that is relevant to the formulation of 
general relativity. (2) A new feature that arises
for non-positive-definite metrics is the presence of {\it null
vectors}, i.e., non-zero vectors whose ``length'' is zero. Attempts to
apply intuition from Riemannian geometry to null vectors and null
surfaces (i.e., surfaces that are everywhere orthogonal to a null
vector) often result in serious errors!

When I teach general relativity at either the undergraduate or
graduate level, I emphasize to the students that one of their main
challenges is to ``unlearn'' some of the fundamental falsehoods about
that nature of space and time that they have been taught to assume
since high school (if not earlier). We have already discussed above
one such key falsehood, namely the notion of absolute
simultaneity. Normally, students taking general relativity have had
some prior exposure to special relativity, and thus they are
aware---at least at some level---of the lack of a notion of absolute
simultaneity in special relativity. However, very few students have
any inkling that, in nature, the points of space and/or the events in
spacetime fail to have any natural vector space structure. Indeed, the
concept of a ``vector'' is normally introduced to students early in
their physics education through the concept of ``position vectors''
representing the points of space! Students are taught that, given the
choice of a point to serve as an ``origin'', it makes sense to add and
scalar multiply points of space. The only significant change
introduced by special relativity is the generalization of this vector
space structure from space to spacetime: In special relativity, the
position vector $\vec{x}$ representing a point of space is replaced by
the ``4-vector'' $x^\mu$ representing an event in spacetime. One can
add and/or scalar multiply 4-vectors in special relativity in exactly
the same way as one adds and/or scalar multiplies ordinary position
vectors in pre-relativity physics.

This situation changes dramatically in general relativity,
since the vector space character of space and/or spacetime depends
crucially on having a flat geometry. In general relativity, it does
not make any more sense to ``add'' two events in spacetime than it
would make sense to try to define a notion of addition of points on
the surface of a potato.

How does one go about giving a precise mathematical description of the
geometry of a spacetime in general relativity---or, for that matter,
of the geometry of a surface of a potato? The notion of a ``distance
function'' between (finitely separated) points can be defined for the
surface of a potato, and, similarly, the notion of a ``spacetime
interval'' could be defined for (finitely separated, but sufficiently
close) events in general relativity, but it would be extremely
cumbersome to base a geometrical description of these entities on such
a notion. A much better idea is to work infinitesimally, using the
idea that, on sufficiently small scales, a curved geometry looks very
nearly flat. These departures from flatness can then be described via
differential calculus. To do so, one begins by introducing the notion
of a tangent vector to describe an infinitesimal displacement about a
point $p$. The collection of all tangent vectors at $p$ can be given
the natural structure of a vector space, but in a curved geometry, a
tangent vector at $p$ cannot naturally be identified with a tangent
vector at a different point $q$. One then uses basic constructions of
linear algebra to define the more general notion of tensors at $p$. A
particularly important example of a tensor field (i.e., a tensor
defined at all points $p$) is a {\it metric}, which is simply a (not
necessarily positive definite) inner product on tangent vectors (see
below). When a metric (of any type) is present, it gives rise to a
natural notion of differentiation of tensor fields. This notion of
differentiation allows one to define the notion of a geodesic (as a
curve that is ``as straight as possible'') and curvature---which can
be defined in terms of the failure of initially parallel geodesics to
remain parallel, or, more directly, in terms of the failure of
successive derivatives of tensor fields to commute.

Let me now explain in more detail what is actually needed in order to
introduce the above basic concepts of differential geometry in a
mathematically precise manner. First, one needs a mathematically
precise notion of the ``set of points'' that constitute spacetime (or
that constitute a surface in ordinary geometry). The appropriate
notion is that of a {\it manifold}, which is a set that locally
``looks like'' $R^n$ with respect to differentiability properties, but
has no metrical or other structure. The points of an $n$-dimensional
manifold can thereby be labeled locally by coordinates
$(x^1,...,x^n)$, but these coordinate labels are arbitrary and could
equally well be replaced by any other coordinate labels $({x'}^1,...,
{x'}^n)$ that are related to $(x^1,...,x^n)$ in a smooth, nonsingular
manner. A precise definition of an $n$-dimensional manifold can be
given as a set that can be covered by local coordinate systems that
satisfy suitable compatibiliy conditions in the overlap regions.

Unfortunately, it is not as easy as one might think to give a
mathematically precise notion of a ``tangent vector''. The most
elegant and mathematically clear way of proceeding is to define a
tangent vector to be a ``derivation'' (i.e., directional derivative
operator) acting on functions; derivations can be defined
axiomatically in a simple manner. This definition has the virtue of
stating clearly what a tangent vector is, without introducing
extraneous concepts like coordinate bases. Essentially all modern
mathematics books define tangent vectors in this way. However, most
students do not find this definition to be particularly intuitive.

A more intuitive way of proceeding is to consider a curve, which can
locally described by giving the coordinates $x^\mu (t)$ of the point
on the curve as a function of the curve parameter $t$. One can
identify the tangent to the curve at the point $x^\mu (t)$ with the
collection of $n$ numbers, $(dx^1/dt,...,dx^n/dt)$, at the point on
the curve labeled by $t$. The coordinate lines themselves are curves,
and the tangent to the $\mu$th coordinate line would be identified
with the numbers $(0,...,0,1,0,...0)$, where the ``$1$'' is in the
$\mu$th place. One may therefore view the tangents to the coordinate
lines at each point as comprising a basis for the ``tangent vectors''
at that point. For an arbitrary curve $x^\mu(t)$, one then may view
$(dx^1/dt,...,dx^n/dt)$ as the components of the tangent to this curve
in this coordinate basis. Of course, if we chose a different
coordinate system, the components of the tangent to this curve would
``transform'' by a formula known as the ``vector transformation law'',
which is easily derived from the chain rule.

A somewhat more direct way of proceeding in accord with the previous
paragraph is to define a tangent vector at a point to be a collection
of $n$ numbers associated with a coordinate system that transforms via
the vector transformation law under a change of coordinates. This
approach allows one to define a tangent vector in one sentence and
thereby move on quickly to other topics.  This definition can be found
in most mathematics books written prior to the mid-20th century as
well as in most treatments of general relativity written by
physicists.  However, it is not particularly intuitive. Furthermore,
by tying the notion of a tangent vector to the presence of a
coordinate system, it makes it extremely difficult for students to
think about tangent vectors (and tensors---see below) in a
geometrical, coordinate independent way.

After tangent vectors have been introduced, the next step is to define
tensors of arbitrary rank. This is done by a standard construction in
linear algebra. Linear algebra is quite ``easy'' compared with many
other mathematical topics, and students taking a general relativity
class will normally have had a course in linear algebra and/or
considerable exposure to it. Unfortunately, however, the way students
are normally taught linear algebra does not mesh properly with what is
needed for general relativity.  The problem is that in the context in
which students have been exposed to linear algebra, a (positive
definite) inner product is normally present. One then normally works
with the components of tensors in an orthonormal basis. One thereby
effectively ``hides'' the role played by the inner product in various
constructions. One also hides the major distinction between vectors and
dual vectors (see below). In general relativity, the key
``unknown variable'' that one wishes to solve for is the metric of
spacetime, which, as already mentioned above, is simply a
(non-positive definite) inner product on tangent vectors. It is
therefore essential that the all of the basic linear algebra
constructions be done without assuming an inner product, so that the
role of the metric in all subsequent constructions is completely
explicit.

To proceed, given a finite dimensional 
vector space, $V$---which, in the case of interest
for us, would be the tangent space at a point $p$ of spacetime---we
define its dual space, $V^*$, to be the collection of linear maps from
$V$ into $\bf{R}$. It follows that $V^*$ is a vector space of dimension
equal to $V$, but, in the absence of an inner product, there is no
natural way of identifying $V$ and $V^*$. However, given a basis of
$V$, there is a natural corresponding basis of $V^*$. Since $V^*$ is a
vector space, we also can take its dual, thereby producing the
``double dual'', $V^{**}$, of $V$. It is not difficult to show explicitly
that there is a natural way of identifying $V^{**}$ with $V$. 

With this established, a {\em tensor of type $(k,l)$} can then be
defined as a multilinear map taking $k$ copies of $V^*$ and $l$ copies
of $V$ into $R$. On account of the isomorphism between $V$ and $V^*$,
tensors of a given type may be viewed in other equivalent ways. For
example, tensors of type $(1,1)$ are isomorphic to the vector space of
linear maps from $V$ to $V$ and also are isomorphic to the linear maps
from $V^*$ to $V^*$. There are two basic operations that can be
performed on tensors: contraction and taking outer products. All
familiar operations can be expressed in terms of these; for example,
the composition of two linear maps can be expressed in terms of the
outer product of the corresponding tensors followed by a contraction.

All of the assertions of the preceding two paragraphs are entirely
straightforward to establish. However, most students
are not used to distinguishing between between vectors and dual
vectors. Indeed, in the familiar context where one has a positive
definite metric, not only can $V$ and $V^*$ be identified, but the
components of a vector in an orthonomal basis are equal to the
components of the corresponding dual vector in the corresponding dual
basis. Students feel that they ``know'' linear algebra, and they
become bored and impatient if one takes the time to carefully explain
the above ideas.  After all, they took the course to learn about
Einstein's revolutionary ideas about space, time, and gravity, not to
learn why a vector space is isomorphic to its double dual. But if one
doesn't carefully explain the above ideas, the students are guaranteed
to become quite confused at a later stage. In 30 years of teaching
general relativity at the graduate level, I have not found a
satisfactory solution to this problem, and I have always found the
discussion of tensors to be the ``low point'' of the course.

Many treatments of general relativity effectively bypass the above
treatment of tensors by working only with the components of tensors in
bases associated with coordinate systems. Given the ``transformation
law'' for components of tangent vectors under a change of coordinates,
the corresponding transformation law for the components of dual
vectors can be obtained, and the more general ``tensor transformation
law'' for a tensor of type $(k,l)$ can be derived. One can then {\it
define} a tensor of type $(k,l)$ on an $n$-dimensional manifold to be
a collection of $n^{k+l}$ numbers associated with a coordinate system
that transform via the tensor transformation law under a change of
coordinates. This approach is taken in many mathematics books written
prior to the mid-20th century and in many current treatments of general
relativity. It has the advantage that one can then quickly move on to
other topics without spending much time talking about tensors. However, it
has the obvious disadvantage that although students may still be
trained to use tensors correctly in calculations, they usually end
up having absolutely no understanding of what they are.

A {\it metric}, $g$, on a vector space $V$ can now be defined as a
tensor of type $(0,2)$ that is nondegenerate in the sense that the
only $v \in V$ satisfying $g(v,w)=0$ for all $w \in V$ is $v=0$. A
metric is then seen to be equivalent to the specification of an
isomorphism between $V$ and $V^*$. If the metric is positive definite,
it is called {\it Riemannian}, whereas if it is negative\footnote{My
sign convention on the definition of Lorentian metrics corresponds to
that used by most general relativists; however, most particle
physicists use the opposite sign convention, i.e., they take a
Lorentzian metric to be positive definite on a one-dimensional
subspace and negative definite on the orthogonal complement of this
subspace.} definite on a one-dimensional subspace and positive
definite on the orthogonal complement of this subspace, it is called
{\it Lorentzian}. Riemannian metrics describe ordinary curved
geometries (like the surface of a potato), whereas curved spacetimes
in general relativity are described by Lorentzian metrics.

During the past half-century, a major cultural divide has opened up
between mathematicians and physicists with regard to the notation used
for tensors. The traditional notation---which is still used by most
physicists---is to denote a tensor, $T$, of type $(k,l)$ by the
collection of its components ${T^{\mu_1 ...\mu_k}}_{\nu_1 ...\nu_l}$,
where the ``up'' indices correspond to vector indices, and the
``down'' indices correspond to dual vector indices. This notation has
the advantage that basic operations on tensors---like taking outer
products or performing contractions---are expressed in a clear and
explicit way. The isomorphism between vectors and dual vectors that is
provided by the presence of a metric can also be nicely incorporated
into this notation by using the metric to ``raise and lower
indices''. However, the notation effectively forces one to think of a
tensor as a collection of components rather than an object with
legitimate status in its own right that does not require the
introduction of a basis. In reaction to this, essentially all modern
mathematics books adopt an ``index free'' notation for tensors. This
notation makes manifest the proper basis/coordinate independent status
of tensors, but it makes it extremely cumbersome to denote even a
moderately complicated series of operations. In my view, an excellent
compromise is to employ an ``abstract index notation'', which mirrors
the component notation, but where a symbol like ${T^{\mu_1
...\mu_k}}_{\nu_1 ...\nu_l}$ would now stand for the tensor itself,
not its components.

After tensors over an arbitrary vector space have been introduced, one
can return to the manifold context and define a {\em tensor field of
type $(k,l)$} to be an assignment of a tensor of type $(k,l)$ over the
tangent space of each point of the manifold.  The next key step is to
formulate a notion of differentiation of tensor fields. The notion of
differentiation of tensor fields is nontrivial because on a manifold
$M$, there is no natural way of identifying the tangent space at a
point $p$ with the tangent space at a different point $q$, so one
cannot simply take the difference between the tensors at $p$ and $q$
and then take the limit as $q$ approaches $p$. In fact, if we had
no additional structure present beyond that of a manifold, there would be no
unique notion of differentiation; rather there would be a whole class of
possible ways of defining the derivative of tensor fields. These can
be described directly by providing axioms for a notion of a derivative
operator, or, equivalently, it can be done by introducing a notion of
``parallel transport'' along a curve. In mathematical treatments, the
notion of parallel transport is usually introduced in the more general
context of a connection on a fiber bundle. The general notions of
fiber bundles and connections have many important applications in
mathematics and physics (in particular, to the description of gauge
theories), but it would normally require far too extensive a
mathematical excursion to include a general discussion of these topics
in a general relativity course, even at the graduate level.

Although there is no unique notion of differentiation of tensors in a
completely general context, when a metric is present a unique notion
of differentiation is picked out by imposing the additional requirement
that the derivative of the metric must be zero. In Euclidean geometry
(or in special relativity), this notion of differentiation of tensors
corresponds to the partial differentiation of the components of the
tensors in Cartesian coordinates (or in global inertial
coordinates). However, in non-flat geometries, this notion of
differentiation---referred to as the {\it covariant derivative}---does
not correspond to partial differentiation of the components of tensors
in any coordinate system.

Once differentiation of tensors has been defined, a {\it geodesic} can
be defined as a curve whose tangent is {\it parallel transported} along the
curve, i.e., the covariant derivative of the tangent in the direction
of the tangent vanishes. It is not difficult to show that, in
Riemannian geometry, a curve with given endpoints is a geodesic if and only
if it is an extremum (though not necessarily a minimum) of length with
respect to variations that keep the endpoints fixed. Similarly, in
Lorentzian geometry---i.e, in general relativity---a timelike geodesic
(i.e., a geodesic whose tangent has everywhere negative ``norm'' with
respect to the spacetime metric) can be characterized as an extremum of the
proper time, $\tau$, elapsed along the curve. If the curve is
described in coordinates $x^\mu$ by specifying $x^\mu (t)$, 
then $\tau$ is given by 
\begin{equation} 
\tau = \int_a^b \sqrt{- \sum_{\mu,\nu} g_{\mu\nu} \frac{dx^\mu}{dt} \frac{dx^\nu}{dt}} \,\, dt
\label{tau}
\end{equation}

After the above notions have been introduced, curvature may be defined
by any of the following three equivalent ways: (1) The failure of
successive covariant derivatives on tensor fields to commute; (2) The
failure of parallel transport of a vector around an infinitesimal
closed curve to return the vector to its orginal value; (3) the
failure of initially parallel, infinitesimally nearby geodesics to
remain parallel. Curvature is described by a tensor field of type
$(1,3)$, called the {\it Riemann curvature tensor}. After the Riemann
curvature tensor has been defined, all of the essential mathematical
material needed for the formulation of general relativity is in place.

\section{Teaching General Relativity at the Undergraduate Level}

Fortunately, there are not many other courses that are essential
prerequisites for an undergraduate general relativity course. It is,
of course, necessary that students have some prior exposure to special
relativity, since the conceptual hurdles will be too large for a
student with no prior familiarity with special relativity. However, it
should suffice to have seen special relativity as normally introduced
at the level of first year introductory physics courses. It is
important that students have taken classical mechanics at the
undergraduate level, and thereby have had exposure to ``generalized
coordinates'' and Euler-Lagrange variations. It also is useful (but
not essential) for students to have taken an undergraduate
electromagnetism course, since one should understand what an
electromagnetic wave is before trying to learn what a gravitational
wave is.

Teaching general relativity at the undergraduate level poses major
challenges, particularly if the course is only one semester (or, worse
yet, one quarter) in length. In a one-semester undergraduate course,
there is simply not enough time to introduce and properly explain the
mathematical material described in the previous section. Indeed, even
in a year-long course, it clearly would be inadvisable to ``front
load'' all of this mathematical material; if one did so, there would
not likely be many students left in the course by the time one got to
the interesting physical applications of general relativity. 

Clearly, it makes sense to begin an undergraduate relativity course
with a discussion/review of special relativity, preferably emphasizing
the geometrical point of view described in section 2 above. It also
would make sense to try to explain some of the fundamental ideas and
concepts of general relativity at a qualitative level at the beginning
of the course, as also described in section 2. To proceed further,
however, it is necessary to introduce some of the mathematical material
discussed in section 3. In my view, the minimal amount of mathematical
material needed to teach a respectable undergraduate course would
include (i) A clear explanation that spacetime in general relativity
does {\it not} have the structure of a vector space and that
coordinates, $x^\mu$, are merely labels of events in
spacetime---devoid of any physical significance in their own
right. (ii) The introduction of the notion of a tangent vector to a
curve, as described in section 3 above. (iii) The introduction of the
notion of a spacetime metric as a (Lorentzian) inner product on
tangent vectors, and its use for determining the elapsed proper time,
$\tau$, along a timelike curve (see eq.~(\ref{tau}) above). (iv) The
introduction of the notion of a timelike geodesic as a curve that
extremizes $\tau$. The geodesic equation (for timelike geodesics) can
then be derived using Euler-Lagrange variation\footnote{The geodesic
equation for null geodesics could then be introduced by a limiting
procedure after one has derived the equation for timelike geodesics.}.
It is worth noting that the same relation between symmetries and
conservation laws that one has in Lagrangian mechanics (namely,
Noether's theorem) then automatically applies to geodesics, so in a
spacetime with a sufficiently high degree of symmetry, one can
actually solve the geodesic equation (or, more precisely, ``reduce it
to quadratures'') using only constants of motion.

The above will give students the necessary tools to interpret what a
spacetime metric is and what its physical consequences are, since the
key things one needs to know are (a) how to calculate elapsed time
along arbitrary timelike curves and (b) how to determine the timelike
geodesics (which represent the possible paths of freely falling
particles) and null geodesics (which represent the possible paths of
light rays) in a spacetime. However, they will not have the
necessary tools to understand Einstein's equation, so it will be
impossible to derive any solutions, i.e., the students will have to
accept on faith that the spacetimes studied do indeed arise as
solutions to Einstein's equation.

After the above mathematical material has been presented,
one will be in a good position to discuss the Schwarzschild
solution (representing the exterior gravitational field of a spherical
body) and the Friedmann-Lemaitre-Robertson-Walker (FLRW) solutions
(representing spatially homogeneous and isotropic cosmologies). With
regard to the Schwarzschild solution, one can solve the timelike and
null geodesic equations and thereby derive predictions for the motion
of planets and the bending of light. For the FLRW metrics, one can
derive the general form of a metric having homogeneous and
isotropic symmetry in terms of an unknown ``scale factor'', $a(t)$, and
explain how a change in $a$ with time corresponds to the expansion or
contraction of the universe. Although one cannot, of course, derive
the equations for the scale factor that result from Einstein's
equation, one can simply write these equations down and derive their
cosmological consequences. 

Even in a one semester undergraduate course, there should still be
some time left to discuss some other key topics, such as gravitational
radiation and its detection, the black hole nature of the (extended)
Schwarzschild solution, other topics in the theory of black holes,
and topics in modern cosmology. In a year-long undergraduate course,
one should be able to cover all of these topics and also present the
mathematical material related to curvature, so that Einstein's equation 
may be obtained.

\section{Teaching General Relativity at the Graduate Level}

In contrast to undergraduates, graduate students will not be satisfied
if they are asked to accept a major component of a theory on faith,
particularly if they are not even told in a precise and complete way
what that component is. Thus, one simply cannot teach a graduate
course in general relativity without a full discussion of Einstein's
equation.  Consequently, it is necessary to introduce the mathematical
material needed to define curvature.

When I have taught general relativity at the graduate level, I have
spent the first two weeks with a discussion/review of special
relativity from the geometrical point of view and a
qualitative discussion of the fundamental concepts underlying general
relativity. I have then launched into a complete exposition of all of
the mathematical material described in section 3 above, ending with a
derivation/discussion of Einstein's equation. This mathematical
portion of the course normally occupies approximately 5 weeks. In a
one-semester (or, worse yet, a one-quarter) course, this leaves enough
time only for a ``bare bones'' treatment of the following essential
topics: (i) ``weak field'' properties of general relativity (Newtonian
limit and gravitational radiation), (ii) the FLRW metrics (see above)
and their key properties (cosmological redshift, ``big bang'' origin,
horizons), and (iii) the Schwarzschild solution (planetary motion, the
bending of light, and the black hole nature of the extended
Schwarzschild metric). I believe that a course of this nature provides
students with a solid introduction to general relativity. By providing
the key conceptual ideas and the essential mathematical tools, it
leaves students well prepared to continue on in their study of general
relativity. However, a course of this nature has the serious drawback
that a high percentage of the effort is spent on mathematical
material, and some students are justifiably frustrated with the minimal
discussion of physical applications of the theory.

In a one-semester course, the only way one could add significantly
more discussion of such physically interesting and relevant topics as
gravitational radiation, black holes, relativistic astrophysics, and
cosmology would be to significantly cut down on the time spent on the
mathematical material. If one introduces coordinates at the outset and
works exclusively with the components of tensors in coordinate (or
other) bases, then, as already described above in section 3, one can
bypass much of the mathematical discussion of tensors by defining
tensors via the tensor transformation law. One then can define
differentiation of tensors by introducing the Christoffel symbol as
the ``correction term'' that needs to be added to the ``ordinary
derivative'' so as to produce a tensor expression (i.e., so as to
produce a collection of components that transforms via the tensor
transformation law under coordinate changes). One can then introduce
the Riemann curvature tensor as an object constructed out of the
Christoffel symbol and its ordinary derivative that---rather
magically---can be shown to transform as a tensor. The main price paid
by presenting the mathematical material in this way is a sacrifice of
clarity in explaining the fundamental conceptual basis of general
relativity---particularly its difference from all prior theories with
regard to the nonexistence of any non-dynamical background structure
of spacetime---since this conceptual basis is very difficult to
understand if one does not formulate the theory in a coordinate
independent way. In addition, students will not have the necessary
mathematical tools to advance their study of general relativity to
topics involving ``global methods''---such as the singularity theorems
and the general theory of black holes---where it is essential that the
concepts be formulated in a coordinate independent way. Nevertheless,
by proceeding in this manner, one can easily reduce the time spent on
mathematical material by a factor of 2 or more, thereby allowing
significantly more course time to be spent on physical applications.

\section{Resources} 

Note: $E=$ elementary level/general interest, $I=$ intermediate level,
$A=$ advanced level/specialized material.

\subsection{Resources for introductory discussions of general relativity}

\noindent
{\bf Relativity: The Special and the General Theory, The Masterpiece
Science Edition}, A. Einstein (Pi Press, New York, 2005). This reprint
of one of Einstein's early, non-technical expositions of special and
general relativity contains an introduction by R. Penrose and
commentary by R. Geroch and D. Cassidy. (E)
\medskip

\noindent
{\bf Flat and Curved Space-Times} (second edition), G.F.R. Ellis and
R. Williams (Cambridge University Press, Cambridge, 2000). This book
provides a discussion of special relativity from a geometrical point
of view and an introduction to the basic ideas of general
relativity. (E)
\medskip

\noindent
{\bf General Relativity from A to B}, R. Geroch (University of Chicago
Press, Chicago, 1978). This book presents an excellent introduction to
the basic ideas of general relativity from a thoroughly geometrical point of
view. (E)
\medskip

\noindent
{\bf Gravity from the Ground Up}, B. Schutz (Cambridge University
Press, Cambridge, 2003). This book provides a very readable discussion
of the nature of gravitation in general relativity and its
implications for astrophysics and cosmology. (E)
\medskip

\noindent
{\bf Exploring Black Holes: Introduction to General Relativity},
E.F. Taylor and J.A. Wheeler (Addison Wesley Longman, San Francisco,
2000). This book provides a very physically oriented introduction to
general relativity and black holes. (E)
\medskip

\noindent
{\bf Black Holes and Time Warps: Einstein's Outrageous Legacy},
K.S. Thorne (W.W. Norton, New York, 1994). This book provides a very
well written account of some of the most fascinating ideas and
speculations to arise from general relativity. (E)
\medskip

\noindent
{\bf Space, Time, and Gravity: The Theory of the Big Bang and Black
Holes} (second edition), R.M. Wald (University of Chicago Press,
Chicago, 1992). (E)
\medskip

\noindent
{\bf Was Einstein Right?: Putting General Relativity to the Test}
(second edition) C.M. Will (Basic Books, New York, 1993). This book
provides an excellent account of the observational and experimental
tests of general relativity. (E)

\subsection{Resources for differential geometry}

\noindent
{\bf Geometry of Manifolds}, R.L. Bishop and R.J. Crittenden (American
Mathematical Society, Providence, 2001). This consise book provides an
excellent, high-level account of differential geometry. (A)
\medskip

\noindent
{\bf Tensor Analysis on Manifolds}, R.L. Bishop and S. Goldberg (Dover
Publications, New York, 1987). (I)
\medskip

\noindent
{\bf Riemannian Geometry}, L.P. Eisenhart (Princeton University Press,
Princeton, 1997). This is a reprint of the 1925 classic monograph,
which gives an excellent presentation of the coordinate-based approach
to differential geometry taken by mathematicians prior to the middle
of the 20th century and still used by most physicists today. (I,A)
\medskip

\noindent
{\bf Foundations of Differential Geometry}, volumes 1 and 2,
S. Kobayashi and K. Nomizu (John Wiley and Sons, New York, 1996). This
book is an excellent, high-level reference on differential
geometry. (A)
\medskip

\noindent
{\bf Riemannian Manifolds : An Introduction to Curvature},
J.H. Lee (Springer-Verlag, New York, 1997). (I)
\medskip

\noindent
{\bf Tensors, Differential Forms, and Variational Principles},
D. Lovelock and H. Rund (Dover Publications, New York, 1989). (I)
\medskip

\noindent
{\bf A Comprehensive Introduction to Differential Geometry}, volumes
1-5, third edition, M. Spivak (Publish or Perish Inc., Houston,
1999). (I)
\medskip

\noindent
{\bf Tensors and Manifolds: With Applications to Mechanics and
Relativity}, R.H. Wasserman (Oxford University Press, Oxford,
1992). This book provides an extremely clear an complete treatment of
the basic definitions, constructions, and results associated with
tensor fields on manifolds. (I)
\medskip

\subsection{Undergraduate level texts}

\noindent
{\bf Gravity: An Introduction to Einstein's General Relativity},
J.B. Hartle (Addison Wesley, San Francisco, 2003). The philosophy on
teaching general relativity to undergraduates expounded in this
Resouce Letter is adopted directly from the approach taken by Hartle
in this text. (I)
\medskip

\noindent
{\bf General Relativity: A Geometric Approach}, M. Ludvigsen (Cambridge
University Press, Cambridge, 1999). (I)
\medskip

\noindent
{\bf Relativity: Special, General, and Cosmological} W. Rindler
(Oxford University Press, Oxford, 2001). (I)
\medskip

\noindent
{\bf A First Course in General Relativity}, B. Schutz (Cambridge
University Press, Cambridge, 1985). (I)
\medskip

\noindent
{\bf Relativity : An Introduction to Special and General Relativity}
third edition, H. Stephani (Cambridge University Press, Cambridge,
2004). (I)
\medskip

\subsection{Graduate level texts/monographs}

\noindent
{\bf Spacetime and Geometry: An Introduction to General Relativity},
S. Carroll (Addison Wesley, San Francisco, 2004). This book provides a
well written, pedagogically oriented introduction to general
relativity. (I)
\medskip

\noindent
{\bf The Large Scale Structure of Space-time}, S.W. Hawking and
G.F.R. Ellis (Cambridge University Press, Cambridge, 1973). This book
is true masterpiece, containing a complete exposition of the key
global results in general relativity, including the singularity
theorems and the theory of black holes. It is not light reading,
however. (A)
\medskip

\noindent
{\bf Relativity on Curved Manifolds}, F. de Felice and C.J.S. Clarke
(Cambridge University Press, Cambridge, 1990). (I,A)
\medskip

\noindent
{\bf The Classical Theory of Fields}, L.D. Landau and E.M. Lifshitz,
(Elsevier, Amsterdam, 1997). This very clear and consise discussion of
general relativity from a coordinate-based point of view occupies only
about 150 pages of this book.  (I,A)
\medskip

\noindent
{\bf Gravitation}, K.S. Thorne, C.W. Misner, and J.A. Wheeler
(W.H. Freeman, San Francisco, 1973). This book, which remains very
widely used, was the first text to present general relativity from a
modern point of view. It places a strong emphasis on the physical content
of the theory. (I,A)
\medskip

\noindent
{\bf Advanced General Relativity}, J. Stewart, (Cambridge Monographs on
Mathematical Physics, Cambridge University Press, Cambridge, 1991). (A)
\medskip

\noindent
{\bf General Relativity}, R.M. Wald (University of Chicago Press,
Chicago, 1984). (I,A)
\medskip

\noindent
{\bf Gravitation and Cosmology : Principles and Applications of the
General Theory of Relativity}, S. Weinberg (Wiley, New York,
1972). This book takes and anti-geometrical approach and some of the
discussion of cosmology is out of date, but it remains one of the best
references for providing the details of calculations arising in the
applications of general relativity, such as to physical processes
occurring in the early universe.(I,A)

\end{document}